\documentclass{ws-procs9x6}

\newcommand{\measure}[1]{\mathrm{d}#1\,}
\def\sumint{\hbox{$\sum$}\!\!\!\!\!\!\int}
\def\sumdiff{\hbox{$\Delta$}\!\!\!\!\!\!\int}
\def\smallsumdiff{\scalebox{0.75}{\hbox{$\Delta$}}\!\!\!\!\!\int}

\begin{document}

\title{Thermodynamics of the 1+1-dimensional nonlinear sigma model
through next-to-leading order in $1/N$} 

\author{Harmen J. Warringa}

\address{Department of Physics and Astronomy,
Vrije Universiteit, \\ 
De Boelelaan 1081, 1081 HV Amsterdam, 
The Netherlands \\
E-mail: harmen@nat.vu.nl}

\maketitle

\abstracts { We discuss the thermodynamics of the $O(N)$ nonlinear
sigma model in 1+1 dimensions. In particular we investigate the NLO
$1/N$ correction to the 1PI finite temperature effective potential
expressed in terms of an auxiliary field. The effective potential
contains temperature-dependent divergences which cannot be
renormalized properly. We argue that this problem vanishes at the
minimum of the effective potential. Therefore physical quantities like
the pressure are well defined and can be renormalized in a
temperature-independent way. We give a general argument for the
occurrence of temperature-dependent divergences outside the
minimum. We present calculations of the pressure and show that $1/N$
is a good expansion. It turns out that the pressure normalized to that
at infinite temperature is $N$-independent like the flavor
independence of the same quantity in QCD.  }

\section{Introduction}
It is interesting to study the thermodynamics of the $O(N)$ nonlinear
sigma model (NLSM) in $1+1$ dimensions in a $1/N$ expansion. It allows
one to learn about renormalization at finite temperature ($T$) in a
non-perturbative approach. In general one expects that temperature
does not influence the ultraviolet behavior of the theory. Hence if
the theory is renormalized at $T=0$, it should also be renormalized at
finite $T$. However, this statement is not true for the effective
potential as a function of the vacuum expectation value (vev) of an
auxiliary field as we will see. The $1+1$-dimensional NLSM can also be
used as a toy model to study QCD. This is because both theories are
asymptotically free. Unlike in large $N_c$ QCD one can calculate the
pressure of the NLSM through NLO in $1/N$. By performing this
calculation we were able to test the $1/N$ expansion at finite
temperature.

The $O(N)$ symmetric NLSM is defined by the Lagrangian density
\begin{equation}
  \mathcal{L} = \frac{1}{2} \partial_\mu 
  \phi_i \partial^\mu \phi_i
   \;,\;\;\;\;\;\;\;
   \phi_i(x) \phi_i(x) = \frac{N}{g_b^2} \;,
  \;\;\;\;\;\; i = 1,\,\ldots,\,N \;,
\end{equation}
where $g_b$ is the bare coupling constant and $\phi_i$ are scalar
fields which are constrained to live on a hypersphere. The NLSM is
renormalizable because $g_b$ is dimensionless. It is asymptotically
free and contains infrared renormalon ambiguities \cite{David}. The
scalar fields become massive due to the interactions. The mass is
non-analytic in the coupling constant.

The effective potential of the NLSM has been studied through NLO in
the $1/N$ expansion at $T=0$
\cite{Root,Novikov,Rim,Biscari,Flyvbjerg}. The LO contribution at
finite $T$ was also investigated before \cite{Dine}. We studied the
NLO correction at finite $T$ \cite{abw}.

\section{The effective potential}
\label{sec:oneovernexp}
A way to implement the constraint on the $\phi$ fields is by
introducing an auxiliary Lagrange multiplier field which we will
denote by $\alpha$.  The action then still remains quadratic in the
$\phi$ fields. If one integrates over the $\phi$ fields one finds the 
partition function of the NLSM which is
\begin{equation}
   \mathcal{Z} =  \int \mathcal{D} \alpha
    \exp \left \{
-\frac{N}{2} \mathrm{Tr} \log[ -\partial^2 - i \alpha(x)]
     - \frac{i N}{2 g_b^2} \int_0^{1/T} \!\!\!\!
  \measure{\tau} \int \measure{x}
       \alpha(x) \right \} \;.
  \label{eq:partf}
\end{equation}
The effective potential can be calculated from the partition function
by expanding the $\alpha$ field around its vev. One can prove that the
vev of the $\alpha$ field is purely imaginary. Therefore we write
$\alpha = i m^2 + \tilde \alpha / \sqrt{N}$. By expanding the
$\mathrm{Tr} \log$ term of Eq.\ (\ref{eq:partf}) around $i m^2$, one
sees that $\tilde \alpha$ n-point vertices are suppressed with respect
to the propagator in the large-$N$ limit. Hence, the effective
potential can be calculated in a $1/N$ expansion.  The LO contribution
stems from the classical action of the $\alpha$ field. The NLO
correction can be found by performing the Gaussian integral over the
quantum fluctuations $\tilde \alpha$.  One finds $\mathcal{V}(m^2) = N
\mathcal{V}_\mathrm{LO}(m^2) + \mathcal{V}_\mathrm{NLO}(m^2)$ where
\begin{eqnarray}
 \mathcal{V}_\mathrm{LO} =
  \frac{m^2}{2 g_b^2} -\frac{1}{2} \sumint_P \log(P^2 + m^2)  \;,
  \;\;\;\;\;\;\;
 \mathcal{V}_\mathrm{NLO} &=&
 -\frac{1}{2} \sumint_P \log \Pi(P, m) \;.
\end{eqnarray}
The Fourier transform of the inverse $\tilde \alpha$ propagator is
given by
\begin{equation}
  \Pi(P,m) = \frac{1}{2} \sumint_Q \frac{1}{(P+Q)^2 + m^2}
  \frac{1}{Q^2 + m^2} \;.
\end{equation}
For convenience we use the following notation, $P = (p_0, p)$ and
\begin{equation}
  \int_P \equiv \int \frac{\mathrm{d^2}p}{\left( 2\pi \right)^2}
  \;,\;\;\;\;\;\;\;
  \sumint_P \equiv T \!\!\! \sum_{p_0 = 2 \pi n T}
  \int \frac{\mathrm{d} p}{2\pi} 
  \;,\;\;\;\;\;\;\;
  \sumdiff_P \equiv \sumint_P - \int_P \;.
\end{equation}

To deduce the required renormalization of the effective potential, we
write $\mathcal{V} = N m^2 / 2 g_b^2 + N D_\mathrm{LO} + N F_\mathrm{LO}
+ D_\mathrm{NLO} + F_\mathrm{NLO}$, where $F_\mathrm{LO}$ and
$F_\mathrm{NLO}$ are finite quantities which have to be calculated
numerically. The terms $D_\mathrm{LO}$ and $D_\mathrm{NLO}$ are
divergent and will be calculated analytically. The theory is infrared
finite, so the divergences will arise from the high-momentum modes. To
regulate the ultraviolet divergences, we introduce a momentum cutoff
$\Lambda$. The quantity $\smallsumdiff_P \log \left[ f(P,m)\right]$ is
finite due to Bose-Einstein suppression. Hence the divergences of the
effective potential stem from $\int_P \log \left[ \lim_{P \gg m,T}
f(P,m) \right]$. In this way one finds $ D_\mathrm{LO} = -m^2 \log
\left(\Lambda^2 /m^2 \right) / 8\pi$. It turns out \cite{abw} that
$\Pi(P,m)$ has an explicit $T$-dependence in the large-momentum
limit. As a result we find
\begin{multline}
   D_\mathrm{NLO} = -\frac{1}{8 \pi} \left[ \Lambda^2 \log \log
   \left( \frac{\Lambda^2}{\bar m^2}\right) - \bar m^2 \mathrm{li}
   \left( \frac{\Lambda^2}{\bar m^2} \right) \right] \\
   + \frac{m^2}{4
   \pi} \left[ \log \left( \frac{\Lambda^2}{m^2} \right) - \log \log 
   \left( \frac{\Lambda^2}{\bar m^2} \right)  \right] \;, 
 \label{eq:nlodivs}
\end{multline}
which is $T$-dependent because $\bar m^2 = m^2 \exp\left[-J_1(\beta m)
\right]$. Here $J_1(x) = 4 \int_0^\infty \mathrm{d} t\,n(\omega_t)
/ \omega_t$ with $\omega_t = \sqrt{t^2 + x^2}$ and $n(x) =
\left[\exp(x)-1 \right]^{-1}$. Hence the quadratic and the
$\mathrm{li}$-divergence have an explicit $T$-dependence. The
logarithmic integral is defined by $\mathrm{li}(x) = \mathcal P \int_0^x
\mathrm{d}t\, 1 / \log t$. 

One possibility to remove the divergences of $\mathcal{V}$ is by
subtracting infinite constants independent of $T$ and $m^2$. This is
allowed because it does not change the shape of $\mathcal{V}$ and
hence not the physics. Clearly this cannot remove the
quadratic divergences, because $\Lambda^2 \left[\log \log(\Lambda^2 /
\bar m^2) - \log \log(\Lambda^2 / \mu^2) \right]$ is still
divergent. The second possibility is to absorb $T$-independent
divergences into $g_b$. This does not solve the problem of the
$\mathrm{li}$-divergence, since it is proportional to $\bar m^2$,
whereas $1/g_b^2$ is proportional to $m^2$ in $\mathcal{V}$. The
$m^2\log \log(\Lambda^2 / \bar m^2)$ divergence can however be
renormalized in this way. The two problematic divergences together are
called \cite{Biscari} ``perturbative tail'' and do not arise in
dimensional regularization. One can remove these divergences by
subtracting the perturbative tail. This however gives rise to a
$T$-dependent ambiguity due to a thermal infrared renormalon
\cite{abw}.

We therefore conclude that we cannot renormalize $\mathcal{V}$ in a
$T$-independent way. However, in the minimum of $\mathcal{V}$, $\bar
m^2 = \Lambda^2 \exp \left( -4\pi / g_b^2 \right) + \mathcal{O}(1/N)$
\cite{abw}. Furthermore one can show \cite{Root} that one only needs
the LO expression for $m^2$ and $\bar m^2$ to calculate $\mathcal{V}$
in the minimum.  Hence the divergences become $T$-independent in the
minimum and they can be removed by subtracting a vacuum
term. Therefore quantities which are to be calculated in the minimum
are well defined. This holds for example for the pressure
$\mathcal{P}$. A summary of the different ways to deal with
$\mathcal{V}$ is shown in Table \ref{tb:effpot}.

\begin{table}[htb]
\tbl{Different ways to deal with the effective potential and their
problems. The subscript min indicates the minimum of the effective
potential.}
{\begin{tabular}{c|l}
quantity & problem \\
\hline \hline
$\mathcal{V}(m^2)$ & $T$-dependent divergences \\
$\mathcal{V}(m^2)- \mathcal{V}(m^2)_{T=0}$ & changes 
shape $\mathcal{V}$ and
$T$-dependent divergences \\
$\mathcal{V}(m^2) - \mathcal{V}_{\mathrm{min}, T=0}$ & 
  $T$-dependent divergences
\\
$\mathcal{V}(m^2) 
 - \mathrm{pert.\;tail}$ & $T$-dependent infrared renormalon 
  ambiguity \\
$\mathcal{P} = \mathcal{V}_{\mathrm{min}} 
- \mathcal{V}_{\mathrm{min},
T=0}$ & no problem 
\end{tabular}}
\label{tb:effpot}
\end{table}

\section{Temperature-dependent divergences}
In this section we will argue in a more general way why it is possible
to have $T$-dependent divergences outside the minimum of the effective
potential. These $T$-dependent divergences are not specific for the
NLSM but also occur in the linear sigma model in $d=3+1$
\cite{abw2}. We add a space-time independent source $J$ for $\alpha$
and define the action in the presence of $J$ as
\begin{equation}
  S[J] = \frac{N}{2} 
  \mathrm{Tr} \log[ -\partial^2 - i \alpha(x)]
     + i \left( \frac{N}{2 g_b^2} + J \right) \int_0^{1/T} \!\!\!\!
  \measure{\tau} \int \measure{x}
       \alpha(x) \;.
\end{equation}
The quantity $Z[J]_T$ is the partition function of the theory with
action $S[J]$ at temperature $T$. If we assume that $\log
Z[0]_T$ does not contain $T$-dependent divergences this also holds for
$\log Z[J]_T$ as long as $J$ is $T$-independent. This is because any
$J$ can be incorporated in $1 / g_b^2$ by a redefinition. The
effective potential in the minimum is equal to $\log Z[0]_T$. Using
this one can deduce from Eq.\ (\ref{eq:nlodivs}) that the divergences
of $\log Z[0]_T$ depend on $1/g_b^2$. Therefore varying $J$ will
change the divergences of $\log Z[J]_T$.

The effective potential is given by $\mathcal{V}(m^2) = \log Z[J]_T /
\beta V- J m^2$, where $J$ is now the current that gives the $\alpha$
field the vev $i m^2$ at a certain $T$. If one varies $T$, the
implicitly defined $J$ has to change in order to keep the same vev $i
m^2$. Therefore $J$ depends on $T$ and as we argued before varying $J$
changes the divergences. Outside the minimum of the effective
potential $J$ is nonzero. This implies that the effective potential
outside the minimum can contain $T$-dependent divergences.

\section{Pressure}
\begin{figure}[th]
 \scalebox{0.6}{\includegraphics{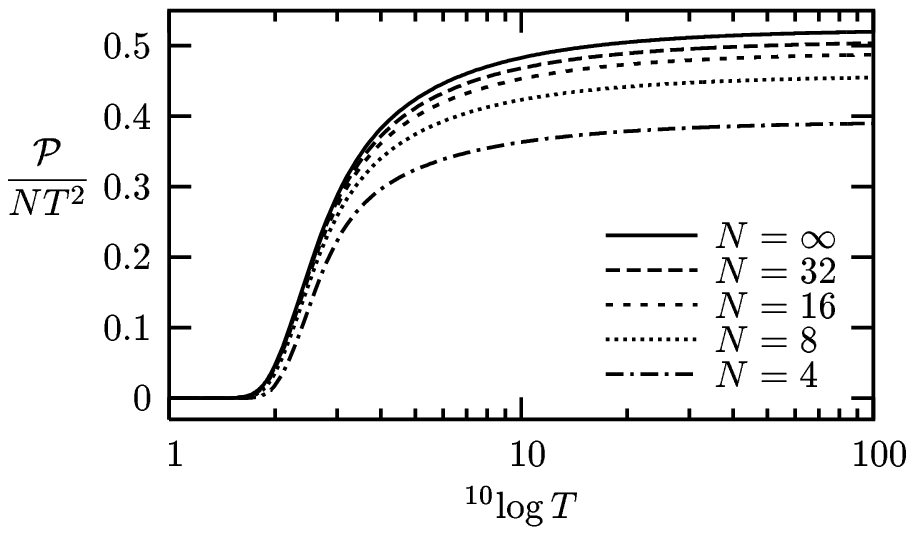}}
 \scalebox{0.6}{\includegraphics{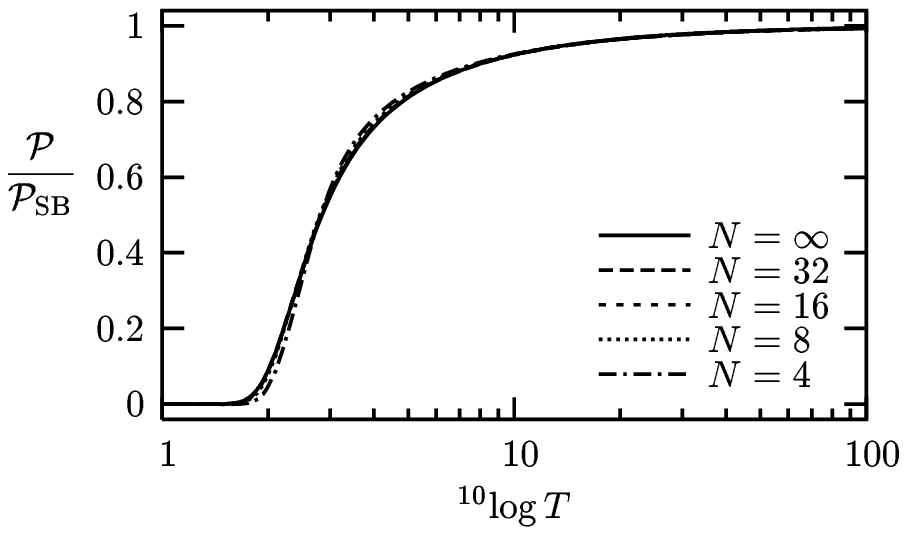}}
\caption{Pressure $\mathcal{P}$ normalized to $N T^2$ (left) 
and normalized to
$\mathcal{P}(T\!=\!\infty)$ (right), as a function of $T$, 
for different values of $N$ with $g^2(\mu = 500) = 10$.}
\label{fig:pressure}
\end{figure}

For the arbitrary choice of $g^2(\mu = 500\;\mathrm{a.u.}) = 10$, we
calculated the pressure $\mathcal{P}$ as function of $T$ for different
values of $N$ \cite{abw}. The results are depicted in \mbox{Fig.\
\ref{fig:pressure}}. With a.u.\ we indicate that all results are in
arbitrary units.

One can see from Fig.\ \ref{fig:pressure} that the NLSM shows a kind
of cross-over transition. This is because at low $T$ the theory is
strongly interacting and hence the degrees of freedom are effectively
frozen. Due to asymptotic freedom, at high $T$ the pressure is
approaching that of a free gas of $N-1$ bosonic degrees of freedom,
i.e$.$, $\mathcal{P} = \mathcal{P_{\mathrm{SB}}} = (N-1) \pi T^2 /
6$. Figure \ref{fig:pressure} also shows that the $1/N$ expansion is a
good expansion for the NLSM. This is because $1/N$ corrections are
really of order $1/N$. The pressure divided by the pressure at
$T=\infty$ is almost independent of $N$. Similar behavior is shown by
QCD lattice calculations with and without dynamical quarks in $d = 3 +
1$ \cite{karsch}, where the same quantity is flavor independent.

\section*{Acknowledgments}
This work has been carried out in collaboration with Jens O. Andersen
and Dani\"el Boer. The author would like to thank the organizers of
the strong and electroweak matter conference for the stimulating
meeting and the opportunity to present this work.


\end{document}